\documentclass[galaxies,accept,pdflatex,10pt,a4paper]{Definitions/mdpi} 

\pubvolume{1}
\issuenum{1}
\articlenumber{0}
\pubyear{2023}
\copyrightyear{2023}
\externaleditor{{Academic Editors:  Lydia Sonia Cidale, Michaela Kraus, María Laura Arias } 
}
\datereceived{14 March 2023} 
\daterevised{1 April 2023} 
\dateaccepted{10 April 2023} 
\datepublished{ } 
\hreflink{https://doi.org/} 

\usepackage[normalem]{ulem}
\newcommand{\degr}{\hbox{$^{\circ}$}}
\newcommand{\arcmin}{\hbox{$^{\prime}$}}
\newcommand{\arcsec}{\hbox{$^{\prime\prime}$}}
\newcommand{\kms}{\,km\,s$^{-1}$}

\newcommand{\Msun}{\ensuremath{\rm M_\odot}}

\newcommand{\HD}{HD\,36030}
\newcommand{\hd}{HD\,36030}
\newcommand{\tess}{\textit{TESS}}
   
\Title{Revealing the Binarity of HD\,36030---One of the Hottest \mbox{Flare Stars}}

\TitleCitation{Revealing the Binarity of HD\,36030---One of the Hottest Flare Stars}

\Author{Olga Maryeva$^{1,}$*\orcidA{}, P{\'e}ter N{\'e}meth$^{1,2}$\orcidC{} and Sergey Karpov$^{3,4}$\orcidB{}} 

\AuthorNames{Olga Maryeva, P{\'e}ter N{\'e}meth, Sergey Karpov} 
\AuthorCitation{Maryeva, O.; N{\'e}meth, P.; Karpov, S.} 

\address{%
$^{1}$ Astronomical Institute of the Czech Academy of Sciences, Fri\v{c}ova 298, 25165 Ond\v{r}ejov, Czech Republic; \\
$^{2}$  {Astroserver.org}, F\H{o} t\'er 1, 8533 Malomsok, Hungary \\
$^{3}$  Institute of Physics of the  Czech Academy of Sciences, Na Slovance 1999/2, 182 00 Prague 8, Czech Republic; \\
$^{4}$  Special Astrophysical Observatory of the Russian Academy of Sciences, 369167 Nizhnii Arkhyz, Russia }

\corres{Correspondence: olga.maryeva@asu.cas.cz}

\abstract{ 
The {\it Kepler} and {\it TESS} space missions significantly expanded our knowledge of what types of stars display flaring activity by recording a vast amount of super-flares from solar-like stars, as well as detecting flares from hotter stars of A-F spectral types.
Currently, we know that flaring occurs in the stars as hot as B-type ones. However, the structures of atmospheres of hot B-A stars crucially differ from the ones of late types, and thus the occurrence of flaring in B-A type stars requires some extension of our theoretical views of flare formation and therefore a detailed study of individual objects. Here we present the results of our spectral and photometric study of \HD, which is a B9\,V star with flares detected by the {\it TESS} satellite. The spectra we acquired suggest that the star is in a binary system with a low-mass secondary component, but the light curve lacks any signs of periodic variability related to orbital motion or surface magnetic fields. Because of that, we argue that the flares originate due to magnetic interaction between the components of the system.
}


\begin{document}

\section{Introduction}

Stars classified as B-type span a very wide range of temperatures (from 25,000 to 11,000~K) and masses (from 50~\Msun\ for B0\,Ia0 down
to 3.5~\Msun\ for B9\,V). 
They usually display the variability related to changes of the photospheric structure, e.g.,~pulsations  of different varieties \citep{McNamara2012}, such as: slowly pulsating B-type stars (SPB) \citep{Waelkens1991,Waelkens1993}, variables of  the $\beta$\,Cephei type \citep{Stankov2005}, variables of the $\alpha$\,Cygni type \citep{Saio2013}, or classical Be (CBe) stars \citep{Rivinius_2013}. 
Less frequently, B-type stars show variability related to  extended atmospheres and outflows, usually observed in peculiar B-type stars, such as B[e]  \citep{Krtickova2018,Kraus2019} or luminous blue variables (LBVs) \citep{Lobel2013,Clark2012}. Quite unexpectedly, some B stars have also been shown to manifest very energetic stellar flares (e.g., \citep{Maryeva2021flare} and references therein)---unpredictable dramatic increases in brightness for a few dozen~minutes.

Such flaring activity is a well-known effect that is often seen both on the Sun and on so-called flare stars.
Energetic flaring events are attributed to the release of the energy stored in coronal magnetic fields in magnetic reconnection events that accelerate the particles (electrons and ions) to high energies, thus heating dense regions of the stellar atmosphere and generating the flare-like emission in a wide range of frequencies, from~radio to gamma-rays \citep{priest_2000,benz2010}.

Although the majority of flare stars were found among objects of late spectral types, primarily cool K-M type red dwarfs \citep{Gershberg1989,Gunther2020}, such outbursts were also detected in some hot B–A type stars, both in the optical range \citep{Schaefer1989,Balona2012, Balona2021} and in X-rays \citep{Schmitt1994,Yanagida2007}. 
In cool stars (F5 and later), the flares are linked with strong magnetic fields produced by a dynamo mechanism working in sufficiently deep outer convection zones \citep{RosnerVaiana1980}. On~the other hand, the~interiors of hot early-type stars are mostly radiative, with~no convective motions to power the dynamo, and~therefore magnetic field production is not expected in them, and~other explanations of the eruptive activity are~needed. 

\textls[-25]{\citet{Pedersen2017} suggested that all flaring hot stars are parts of binary systems and~that the flares occur on unresolved cool components. However,~\citet{SvandaKarlicky2016} and \citet{Jian-Ying2020}} found that the frequency distribution of flare energies for A-type stars is steeper (with more flares having large energies) than the one for stars of later spectral types. It may suggest that the nature of flares on hot stars is indeed different, and~they are not occurring on their colder companions. \mbox{\citet{Balona2012,Balona2021}} studied the variability of flaring B-A type stars and argued that rotational modulation seen in them confirms the presence of strong surface magnetic fields. One more possible mechanism for the flares may be related to magnetic interaction between the components in a close binary system, as was suggested by \citet{Yanagida2007} for the pre-main sequence stars HD\,47777 and HD\,261902. 

Currently, more than one hundred hot flaring B-A type stars are known from observations of the {\it Kepler} and {\it TESS} missions \citep{Balona2012,Maryeva2021flare,Balona2021}. \citet{Doorsselaere2017} demonstrated that about 2.45\% of all B-A stars display detectable flares. However, the~nature of their flares is still elusive. Thus, the~study of these stars, and~their possible binarity, is of utmost~importance.
  
This paper is devoted to the spectral and photometric study of HD\,36030\endnote{RA $05^{h}28^{m}58^{s}\cdot526$; Dec $+03\degr38\arcmin49\arcsec\cdot28$.}, classified as B9\,V \citep{HoukSwift1999} and belonging to the galactic open cluster ASCC\,21 (or [KPR2005]\,21 \citep{KPR2005, Cantat2018}).   For~the first time, HD\,36030 was mentioned as a flare star in \citet{Maryeva2021flare}, where the authors searched for flare stars among confirmed members of galactic open clusters using high-cadence photometry from the {\it TESS} mission. \citet{Maryeva2021flare} found eight B-A type flaring objects; among~them, \HD ~was the hottest. Independently, \citet{Balona2021} and \citet{Yang2023} also selected HD\,36030 as a flare star while searching for flares in {\it TESS} data.
Despite being included in a couple dozens of catalogs, and~being detected as an X-ray source by the Swift X-ray Telescope (XRT) \citep{Evans2020}), to~date \HD ~lacks any specific studies devoted to~it.

In Section~\ref{sec:spectra}, we present new spectral data, their analysis, and~the results of atmospheric modeling using the {\sc Tlusty} code. In~Section~\ref{sec:photometry}, we perform the analysis of photometric data on different time scales, and~in 
Section~\ref{sec:discussion}, we discuss the results, the~parameters of the second component, and the nature of flares. Finally, short conclusions are given in Section~\ref{sec:conclusions}.
 



\section{Spectroscopy of \HD \label{sec:spectra}}
\unskip

\subsection{Observations}
To clarify the nature of \HD ~flaring activity, we initiated its spectral monitoring at the Perek 2 m telescope of  the Ond\v{r}ejov observatory of the Astronomical Institute of the Czech Academy of Sciences. High-resolution spectra of HD\,36030 were obtained  with the Ond\v{r}ejov echelle spectrograph (OES) \citep{OES2004,OES2020} during the autumn of 2021 and spring of 2022 (Table~\ref{tab:log}). The~OES provides a wavelength range  of 3750--9200~\AA~ and a spectral resolving power of 50,000.  The spectra were reduced using a dedicated IDL-based package. After~primary reduction, all spectra were normalized and corrected for barycentric~velocity. 
 
 \begin{table}[H]
 \tablesize{\fontsize{8}{8}\selectfont}
\caption{Log of \HD ~observations at Perek 2 m telescope. \label{tab:log} }
\begin{tabularx}{\textwidth}{lCCC}
\toprule
\textbf{Date}               &   \textbf{ Exposure [s]}       & \textbf{Sp. Range [\AA\AA]} & \textbf{Instrument}    \\    \midrule
       19 Sep 2020         &   3600           &  6250--6740   &    Coud\'e    \\
       24 Oct 2021         &    5200          & 3750--9200    &    OES      \\
       29 Oct 2021         &    5200          & 3750--9200    &    OES      \\
       6 Jan 2022          &    5200          & 3750--9200    &    OES      \\
       2 Feb 2022          &  $3\times5200$   & 3750--9200    &    OES      \\
       10 Mar 2022         &  $2\times5200$   & 3750--9200    &    OES      \\
       11 Mar 2022         &  $2\times5200$   & 3750--9200    &    OES      \\
       18 Mar 2022         &    5200          & 3750--9200    &    OES      \\
\bottomrule
\end{tabularx}
\end{table}
 
Hydrogen lines in the acquired OES spectra have  profiles with broad wings, with~the width of the lines  comparable to an echelle spectral order. For~this reason, for~the spectral normalization in the H$\alpha$ region we used the medium resolution spectrum of HD\,36030 covering a 6250--6740\,\AA\ range acquired using the Coud\'e spectrograph ($R\simeq13000$; \citet{SlechtaSkoda2002}), which was also obtained at the Perek 2-m telescope on 20 September 2020. The~spectrum was reduced using an IDL-based package, in~the same way as described in \citet{Maryeva2022}.  Unfortunately, we have medium-resolution spectra only for the H$\alpha$ region, so we could not perform the same routine for the orders containing other hydrogen lines. For~them, we performed manual continuum drawing, and~thus the normalization in these regions is less~robust.

\subsection{Spectral~Analysis} 

\begin{figure} [H]
{\resizebox*{0.8\columnwidth}{!}{\includegraphics[angle=0]{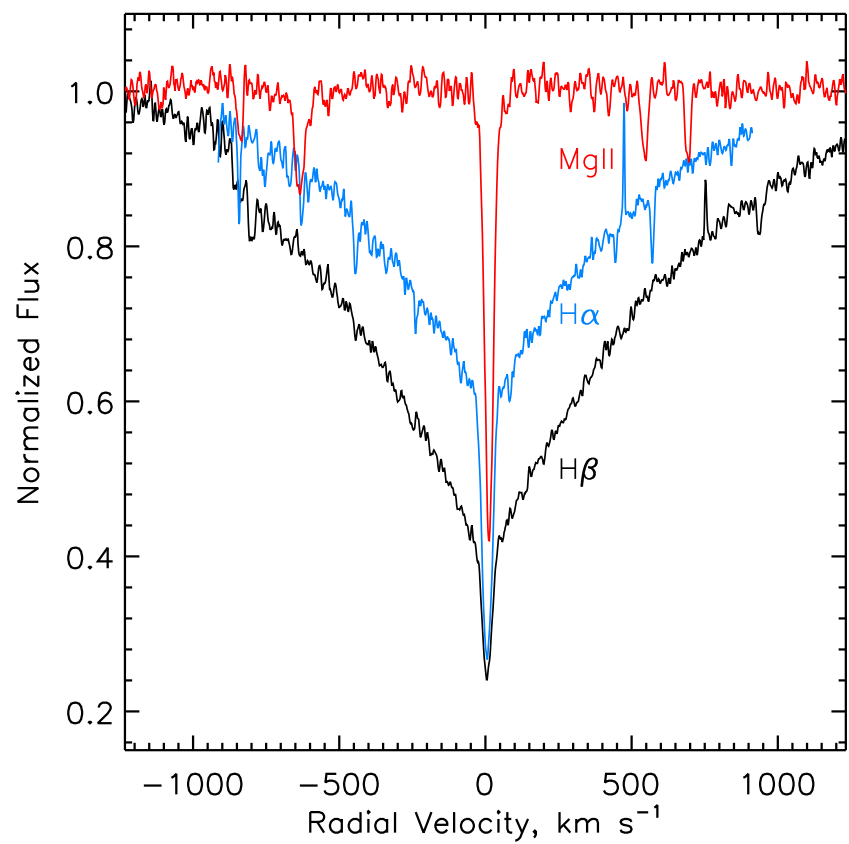}}}
\caption{Comparison	of broad wings hydrogen lines and narrow of Mg\,II $\lambda4481$ observed in the spectrum of HD\,36030.  \label{fig:profiles}}
\end{figure}

\begin{figure} [H]
{\resizebox*{1.0\columnwidth}{!}{\includegraphics[angle=0]{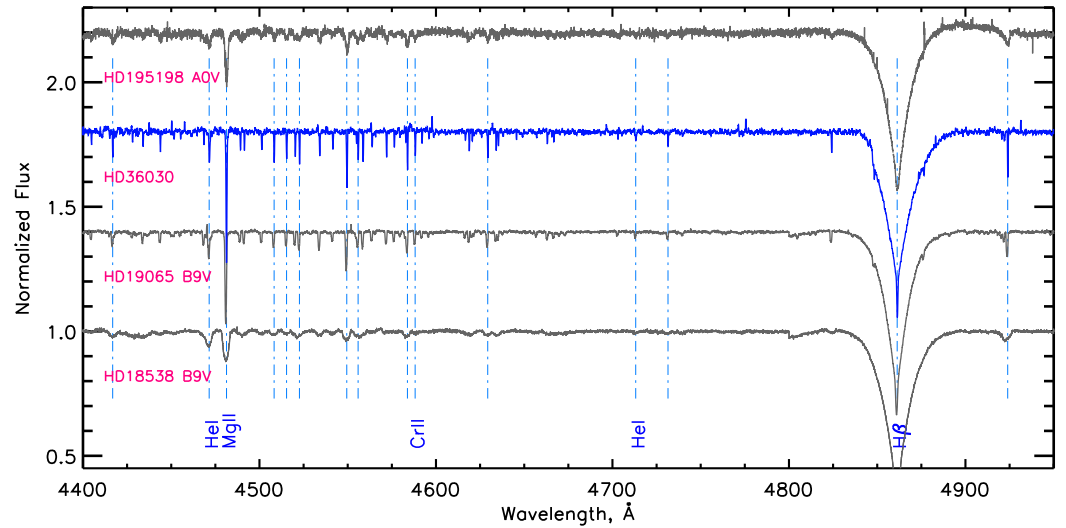}}}
\caption{Comparison of HD\,36030 spectrum with the ones of B9~V and A0~V stars.  Spectra of HD\,18538 and HD\,19065 are taken from IACOB database \citep{IACOB2011,IACOBbase2011,IACOB2015} (spectral resolution is  R = 85,000), whereas the spectrum of HD\,195198 is from~\cite{Prugniel2001} (R = 42,000). Unnamed lines correspond to Fe\,II.
 \label{fig:spectraB9}}
\end{figure}

In the spectrum of \HD, there is a strong Mg\,II $\lambda4481$ line, Ca\,II $\lambda3933,3968$ doublet, Mg\,I $\lambda5167,5172,5183$ triplet, as~well as a number of Fe\,II lines.  All the lines in the spectrum besides the hydrogen lines are narrow. As~it was noted above, the~hydrogen lines show prominent Voigt profiles,  which are a signature of the Stark effect (Figure~\ref{fig:profiles}).  
It is possible to also detect the lines of He\,I $\lambda4471,5876$, which is a piece of evidence that the star is hotter than A0 \citep{SpectralClassification2009}. Figure~\ref{fig:spectraB9} shows a comparison of the HD\,36030's spectrum with ones of B9\,V and A0\,V~stars. 

\begin{figure} [H]
\centerline{\resizebox*{1.0\columnwidth}{!}{\includegraphics[angle=0]{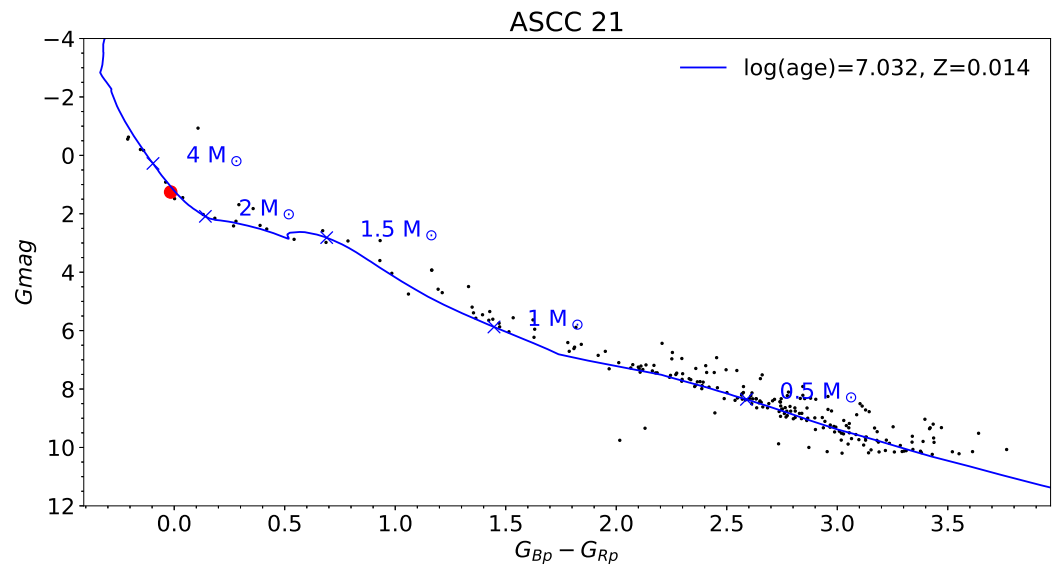}}}
\caption{Hertzsprung--Russell diagram for the ASCC\,21 cluster based on {\it Gaia} photometric data and compared with PARSEC isochrones \citep{PARSEC2014} for the solar metallicity Z = 0.014. Age of the cluster (10.8~Myr) is taken from \citet{Bossini2019}. Red circle shows the position of HD\,36030.
 \label{fig:hr}}
\end{figure}

We estimated the reddening by comparing the photometric data from the Tycho-2 catalogue \citep{tycho2} ($B=8.95\pm0.02$~mag and  $V=8.96\pm0.01$~mag)
with the intrinsic color for a B9 star\endnote{Up to date table of colors and effective temperatures of stars from \citet{Mamajek2013} is available at \url{https://www.pas.rochester.edu/~emamajek/EEM_dwarf_UBVIJHK_colors_Teff.txt}, accessed on March 10, 2023.} ($(B-V)_0=-0.07$~mag; \citet{Mamajek2013}) as $E(B-V)=0.06$~mag. Then, assuming the distance to its host cluster ASCC\,21 to be $d=345.5_{-11.6}^{+12.3}$~pc \citep{Cantat2018}, the~absolute magnitude of \HD ~is $M_V=1.09\pm0.07$~mag. It is fainter than the expected absolute magnitude of a B9 dwarf ($M_{{V} {\rm table}}$ = 0.5~mag from the table of \citet{Mamajek2013}). However, the~location of \HD ~in the Hertzsprung--Russell (HR) diagram for ASCC\,21 cluster (Figure~\ref{fig:hr}) accurately corresponds to the Main Sequence, and~therefore we consider \HD ~to be a B9 dwarf (B9\,V), in~agreement with \citet{HoukSwift1999}.

\begin{figure} [H]
\centerline{\resizebox*{0.95\columnwidth}{!}{\includegraphics[angle=0]{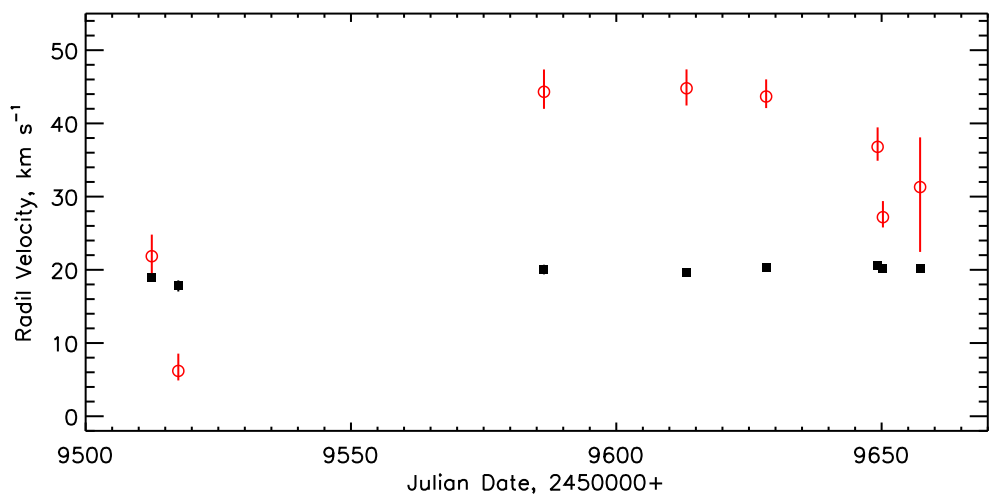}}}
\caption{Radial velocity curve for \hd. Red dots display average values for radial velocity for different lines listed in Table~\ref{tab:rvtab} for a given night. Error bars represent the spread of individual line velocities. For~comparison, black dots show the average velocity of the NaI $\lambda5890, 5896$ interstellar~doublet. 
\label{fig:radvel}}
\end{figure}

Lines in the spectrum of \HD ~clearly display a shift from night to night. We measured radial velocities (RVs) of selected lines (collected in Table~\ref{tab:rvtab}). For~the RV measurements, we used Gaussian profile fittings, and,~for the case of  hydrogen lines, we measured the positions of the narrow central absorptions. As~Table~\ref{tab:rvtab} shows, the~lines of different elements and ions differ in RV. RVs of Mg\,II  $\lambda4481$ and O\,I $\lambda8446$ lines are higher than those of other lines, but~all the lines demonstrate the same pattern of variability. Figure~\ref{fig:radvel} shows average RVs, and~those of the interstellar Na\,I $\lambda5890, 5896$, measured in the same way as a reference. The~scatter of the RV values is from 6 up to 45\kms, and~they clearly show that \HD ~is a binary system. However, we do not see any lines of the secondary component in the spectrum, so we classified \HD ~as a single-lined spectroscopic binary (SB1)~system. 

\vspace{-6pt}

\subsection{Spectral Analysis with {\sc~XTgrid}}
\label{sec:tlusty}

To determine the atmospheric parameters of \HD ~in local thermodynamic equilibrium (LTE), we fitted its OES observations with synthetic spectra calculated from {\sc Tlusty} (v207) model atmospheres \citep{HubenyLanz1995,LanzHubeny2007,HubenyLanz2017}. 
The models include opacities from H, He, C, N, O, Ne, Mg, Si, P, S, and Fe. The spectral analysis was done with a steepest-descent spectral analysis procedure, implemented in the {\sc XTgrid} code \citep{Nemeth2012XTgrid}. 
The procedure is a global fitting method that simultaneously reproduces all line profiles with a single atmosphere model. 
{\sc XTgrid} starts with an input model and, by successive approximations, it calculates new model atmospheres and their corresponding synthetic spectra iteratively in the direction of decreasing $\chi^2$. 
The procedure adjusts the atomic data and microphysics to the changing conditions in the atmosphere as the iterations move across the parameter space. 
Once the fitting procedure reaches the global minimum, statistical errors are calculated by changing each parameter in one dimension until the $\chi^2$ variation corresponds to the 60\% confidence. 
To avoid local minima, the procedure returns to the descent part if a better fit is found during the error analysis.  
Our models evaluated the conditions for convection, but~the convective gradients indicated there are no convective layers in the atmosphere.  
We have made attempts to calculate non-LTE models and evaluate departures from the LTE approximation; however, all such models met numerical instabilities and failed. 
A non-LTE analysis will require the latest version of {\sc Tlusty} (v208) and will be reported in a forthcoming~publication.

\begin{figure} [H]
\resizebox*{0.96\columnwidth}{!}{\hspace{1mm}\includegraphics[angle=0]{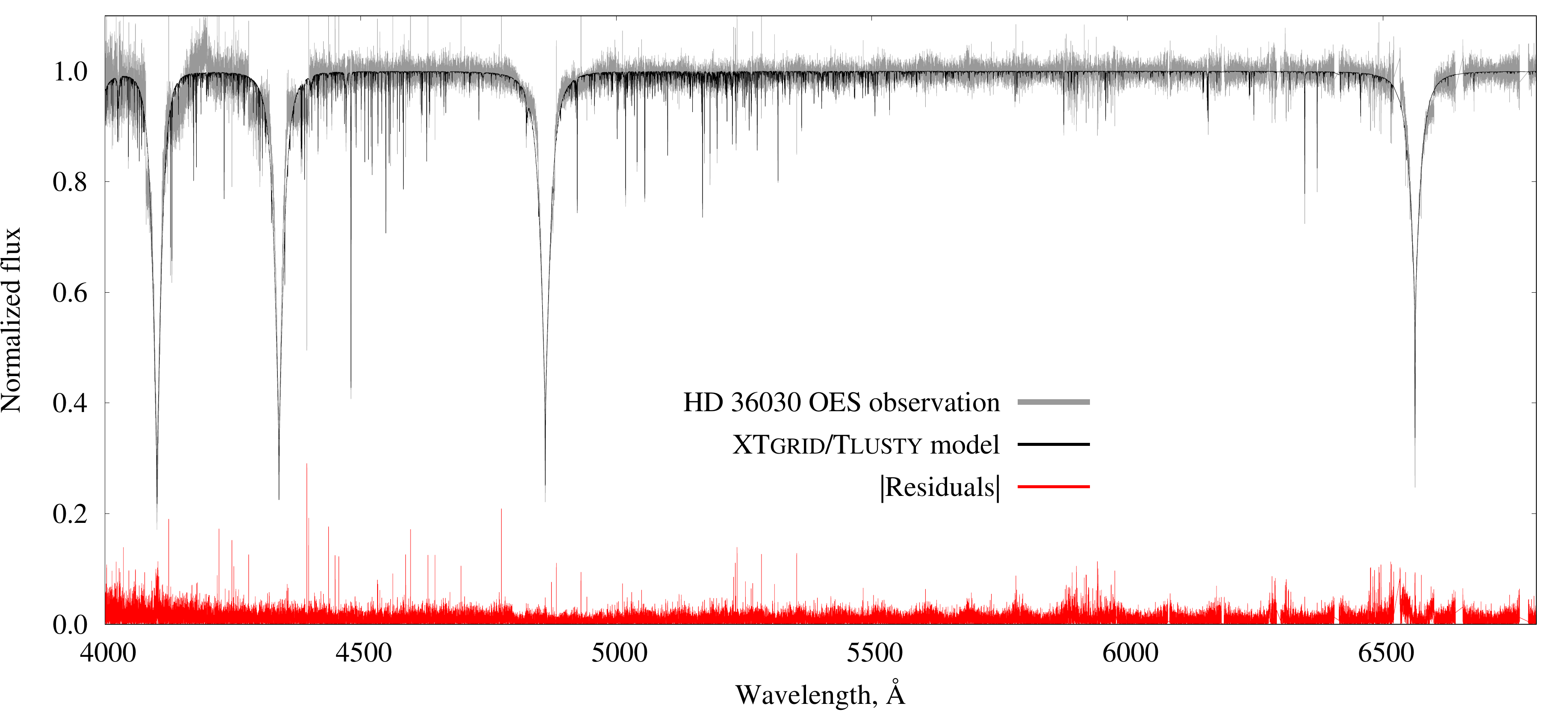}}
\caption{ Normalized spectrum (grey line) of \hd ~taken on 29 October 2021 with OES spectrograph compared with the best-fitting {\sc Tlusty} model (black line). The~absolute values of the residuals between observed and synthetic spectra are shown in red.
 \label{fig:sp_1}}
\end{figure}

\begin{figure} [H]
\resizebox*{1\columnwidth}{!}{\hspace{1mm}\includegraphics[angle=0]{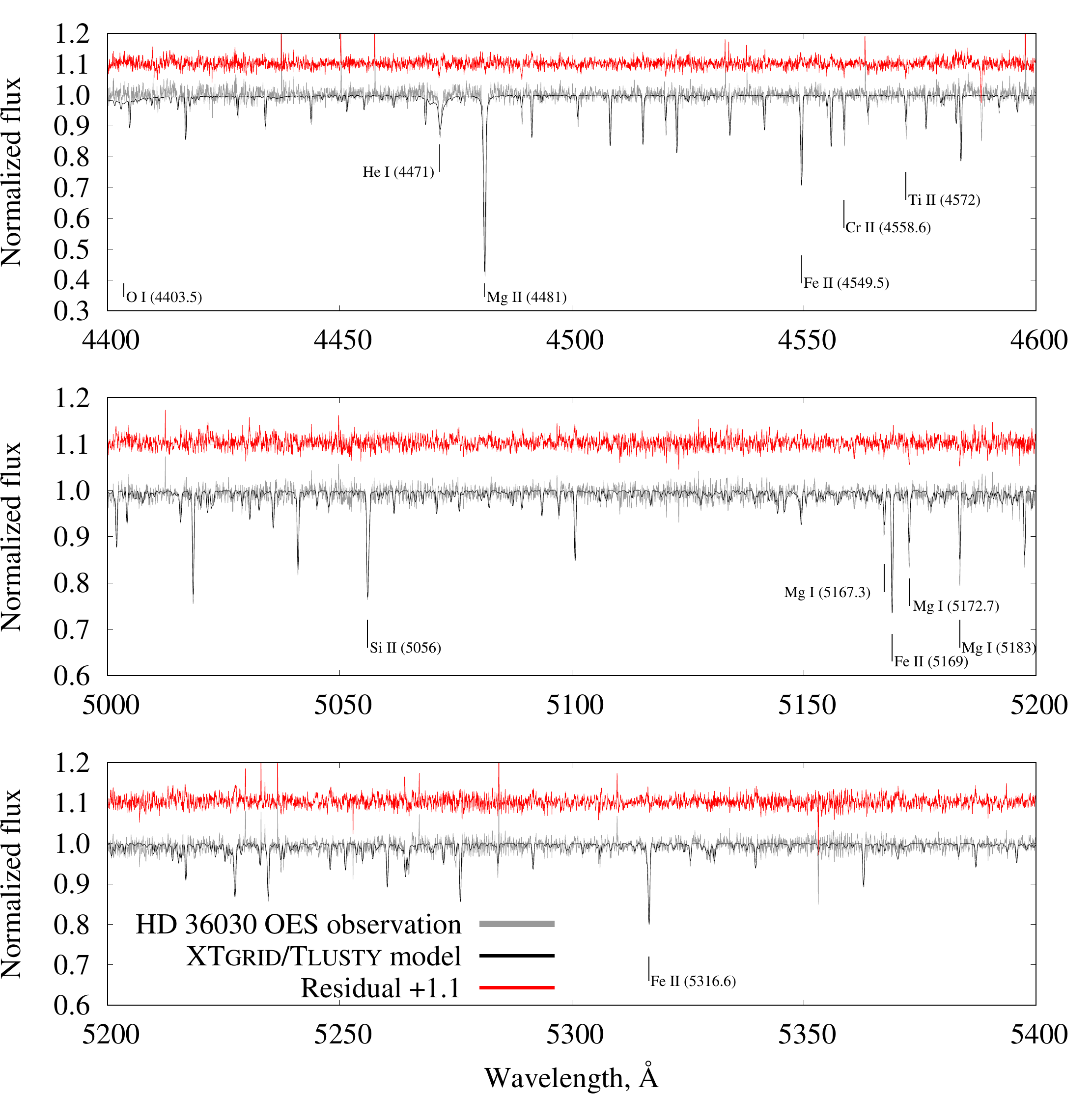}}
\caption{Selected intervals from Figure~\ref{fig:sp_1}.
 \label{fig:sp_2}}
\end{figure}

The spectroscopic parameters obtained from {\sc Tlusty} LTE models are summarized in Table~\ref{tab:tlusty}, and the best fit is shown in Figures~\ref{fig:sp_1} and \ref{fig:sp_2}. 
The large error of $\log{g}$ measurement is mostly due to uncertainties in the continuum normalization of the broad hydrogen lines. Such big uncertainties in $\log{g}$ and $T_{\rm eff}$ give us a broad range of a possible mass of the star, $M=2.1\div9~\Msun$. 
However, the~position of \HD ~in the HR diagram (Figure~\ref{fig:hr}) is consistent with $M_*=3~\Msun$. 

To check the validity of {\sc Tlusty} models, we have repeated the analysis with interpolated {\sc Atlas9} LTE models from the BOSZ spectral library \citep{Bohlin2017}.
The {\sc Atlas9} models confirmed the effective temperature and surface gravity within error bars, but~the BOSZ library was calculated for scaled solar metallicities; therefore, it is not suitable to derive individual element abundances for further~comparisons.

\begin{table}[H]
\tablesize{\fontsize{7}{7}\selectfont}
\caption{Radial velocity measurements for different lines on different nights; Avg is  average values for radial~velocity. The colons after values mark the data with lower accuracy of position determination.}  \label{tab:rvtab}
\begin{adjustwidth}{-\extralength}{0cm}
\begin{tabularx}{\fulllength}{lCCCCCCCC}
\toprule
& \multicolumn{8}{c}{\textbf{Time}\textbf{, JD-2450000}} \\
 & \textbf{9512.463}  & \textbf{9517.464}  &  \textbf{9586.371}  &  \textbf{9613.234}  &   \textbf{9628.234}  & \textbf{9649.242}  & \textbf{9650.239}  &   \textbf{9657.250}    \\

\midrule 
SiII 4128.05       &     21.26  &   7.35     &  44.53      &  45.61      &   42.53      &   37.62    &   27.39    &   35.31        \\
SiII 4130.89       &     21.38  &   6.08     &  42.73      &  42.74      &   42.78      &   36.76    &   25.79    &   38.09:       \\
FeII 4173.45       &     24.71  &   6.17     &  43.38      &  46.39      &   45.07      &   37.99    &   29.35    &   -----     \\
FeII 4233.17       &     19.55  &   5.53     &  43.25      &  45.08      &   43.40      &   36.20    &   27.51    &   30.84        \\
H$\gamma$ 4340.46  &     20.90  &   4.89     &  41.99      &  42.87      &   43.36      &   38.13    &   27.30    &   22.44        \\
MgII 4481.13       &     27.91  &   12.49    &  50.22      &  50.94      &   49.62      &   42.80    &   33.23    &   38.11        \\
FeII 4508.29       &     20.93  &   5.05     &  44.73      &  44.03      &   42.33      &   35.62    &   26.79    &   -----    \\
FeII 4515.34       &     21.28  &   5.71     &  44.66      &  42.45      &   44.85      &   36.23    &   26.38    &   32.61        \\
FeII 4522.63       &     23.46  &   7.39     &  46.31      &  46.10:     &   44.30      &   36.18    &   26.59    &  -----     \\
FeII 4549.47       &     24.81  &   8.55     &  47.36      &  47.38      &   46.01:     &   39.46:   &   29.39    &   36.36        \\
FeII 4555.89       &     22.38  &   5.94     &  43.32      &  45.34      &   43.85      &   37.01    &   26.06    &   31.53        \\
FeII 4583.83       &     22.67  &   6.07     &  43.86      &  44.88      &   43.46      &   36.80    &   27.50    &   33.81        \\
CrII 4588.20       &     20.39  &   6.64     &  45.12      &  44.18      &   44.73      &   36.67    &   26.42    &   30.59        \\
FeII 4629.33       &     20.62  &   5.47     &  44.98      &  43.82      &   42.56      &   36.69    &   26.55    &   29.68        \\
H$\beta$ 4861.32   &     21.70  &   5.63     &  44.51      &  45.16      &   44.30      &   36.38    &   27.23    &   25.98        \\
FeII 4923.92       &     22.13  &   5.95     &  45.11      &  45.21      &   43.15      &   36.97    &   27.57    &   30.14        \\
FeII 5018.44       &     22.00  &   6.52     &  44.92      &  44.76      &   44.00      &   36.79    &   27.83    &   31.05        \\
SiII 5041.03       &     20.91  &   6.32     &  42.70      &  44.29      &   42.67      &   36.25    &   26.98    &   30.38        \\
SiII 5055.98       &     26.04  &   10.63    &  47.98      &  50.36      &   47.87:     &   41.70:   &   32.30:   &   35.37:       \\
FeII 5100.73       &     21.90  &   6.55     &  44.71      &  45.12      &   44.78      &   36.76    &   27.72    &   33.09:      \\
FeII 5169.03       &     21.08  &   5.51     &  43.81      &  44.09      &   43.33      &   37.20    &   26.70    &   30.81        \\
MgI 5172.68        &     21.90  &   7.20     &  45.38      &  45.56      &   44.21      &   36.98    &   28.01    &   31.29        \\
MgI 5183.60        &     22.55  &   5.94     &  44.42      &  45.90      &   44.59      &   36.94    &   28.05    &   31.03        \\
FeII 5197.57       &     22.46  &   5.67     &  43.52      &  43.66      &   42.24      &   36.93    &   26.73    &   30.18        \\
FeII 5275.99       &     23.09  &   6.38     &  42.87      &  46.52:     &   42.09      &   34.90    &   26.02    &   32.96        \\
FeII 5316.62       &     22.64  &   7.90     &  45.74      &  45.18      &   44.90      &   37.40    &   28.14    &   33.12        \\
FeII 5362.87       &     21.55  &   5.46     &  43.33      &  45.05      &   43.59      &   36.20    &   27.05    &   30.11        \\
SiII 6347.09       &     20.44  &   5.18     &  42.55      &  42.56      &   42.20      &   35.66    &   25.80:   &   29.80        \\

SiII 6371.36       &     21.61  &   6.10     &  -----        &  44.44      &   43.67      &   36.53    &   27.15    &   30.10        \\
H$\alpha$ 6562.79  &     21.68  &   6.14     &  46.75      &  45.92      &   43.88      &   37.15    &   27.34    &   31.29        \\
MgII 7877.13       &     -----   &   0.89     &  39.54      &   -----      &  41.64       &   34.30    &   23.38    &   24.77        \\    
MgII 7896.37       &     -----   &   -2.87    &  43.00      &   44.15     &  41.05       &   35.66    &   24.90    &   ------       \\
OI 8446.35         &     28.33  &   12.08    &  49.48      &   48.55     &  49.32       &   40.71    &   32.85    &   34.61        \\
\\
Avg                &$22.40^{+5.94}_{-2.85}$  & $6.72^{+5.77}_{-1.84}$&    $44.58^{+5.63}_{-5.05}$&  $45.15^{+5.79}_{-3.51}$& 
$44.0^{+5.6}_{2.96}$  & $37.14^{5.66+}_{-2.34}$  & $25.52^{+5.71}_{4.14}$  &$31.43^{+6.68}_{8.99}$  \\
\\
NaI 5889.95       &     18.60  & 17.03       &  19.38      &   19.47     &   20.10      &    20.36   &   20.03    &  20.51        \\
NaI 5895.92       &     19.32  & 18.56       &  20.69      &   19.92     &   20.56      &    20.90   &   20.38    &  19.89        \\
\bottomrule
\end{tabularx}
\end{adjustwidth}
\end{table}

\vspace{-8pt}
\unskip   
\vspace{-6pt}
\begin{table}[H]
\caption{Surface parameters of \HD ~from the analysis of the OES spectra.  \label{tab:tlusty} }
\begin{tabularx}{\textwidth}{lCCC}
\toprule
\multicolumn{2}{c}{Parameter}               &   \multicolumn{2}{c}{Value}\\
\midrule
\multicolumn{2}{c}{$T_{\rm eff}$ (K)} & \multicolumn{2}{c}{11,900 $\pm$ 1100}\\
\multicolumn{2}{c}{$\log{g}$ (cm s$^{-2}$)}       & \multicolumn{2}{c}{4.69 $\pm$ 0.15}\\

\multicolumn{2}{c}{v\,$\sin{i}$ (km s$^{-1}$)}       & \multicolumn{2}{c}{15.00 $\pm$ 0.94}\\

\midrule
Abundance   & by number & mass fraction & solar fraction\\
             &$\log{(n{\rm X}/n{\rm H})}$ & &$\log{(\epsilon/\epsilon_\odot)}$\\
\midrule
H (reference)& 1                  &  9.16 $\times$ 10$^{-1}$    & 0      \\
He           & $-$1.75 $\pm$ 0.32 &  6.49 $\times$ 10$^{-2}$    & $-$0.68\\
C            & $<$$-$2.3          &  $<$4.60 $\times$ 10$^{-4}$ & $<$$-$0.8\\
N            & $<$$-$1.9          &  $<$1.36 $\times$ 10$^{-3}$ & $<$$-$0.2\\
O            & $-$3.25 $\pm$ 0.02 &  8.11 $\times$ 10$^{-3}$    & 0.06\\
Ne           & $<$$-$2            &  $<$1.50 $\times$ 10$^{-3}$ & $<$$-$0.03\\
Na           & $-$5.42 $\pm$ 73   &  7.89 $\times$ 10$^{-5}$    & 0.34\\
Mg           & $-$4.39 $\pm$ 0.04 &  9.00 $\times$ 10$^{-4}$    & 0.01\\
Al           & $-$5.89 $\pm$ 0.49 &  3.19 $\times$ 10$^{-5}$    & $-$0.34\\
Si           & $-$4.21 $\pm$ 0.08 &  1.59 $\times$ 10$^{-3}$    & 0.28\\
Ca           & $<$$-$5.48         &  $<$7.3 $\times$ 10$^{-5}$  & $<$$-$0.04\\
Ti           & $-$7.00 $\pm$ 0.11 &  4.38 $\times$ 10$^{-6}$    & 0.05\\
Cr           & $-$6.11 $\pm$ 0.21 &  3.67 $\times$ 10$^{-5}$    & 0.25\\
Mn           & $-$6.94 $\pm$ 0.88 &  5.73 $\times$ 10$^{-6}$    & $-$0.37\\
Fe           & $<$$-$4.13         &  $<$3.74 $\times$ 10$^{-3}$ & $<$0.37\\
Ni           & $<$$-$5            &  $<$5.34 $\times$ 10$^{-4}$ & $<$0.78\\
\bottomrule
\end{tabularx}
\end{table}

\section{Photometry}
\label{sec:photometry}
 
\begin{figure} [H]
    \resizebox*{0.972\columnwidth}{!}{\includegraphics[angle=0]{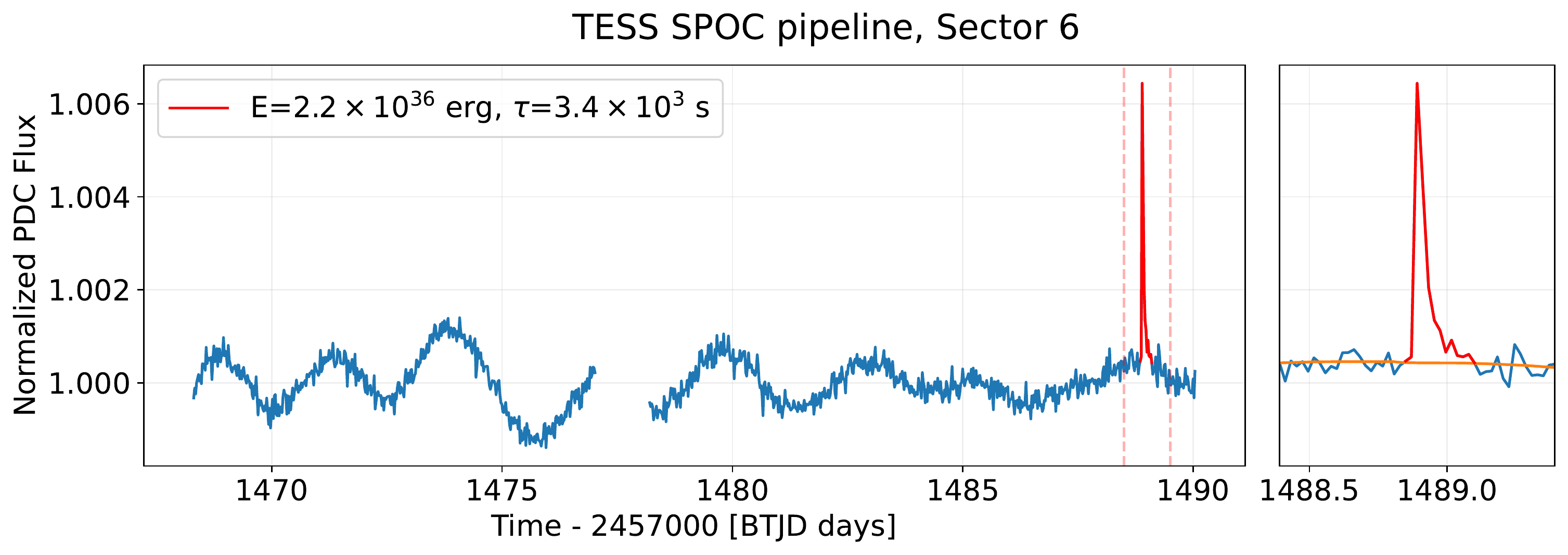}}
    \resizebox*{1.0\columnwidth}{!}{\includegraphics[angle=0]{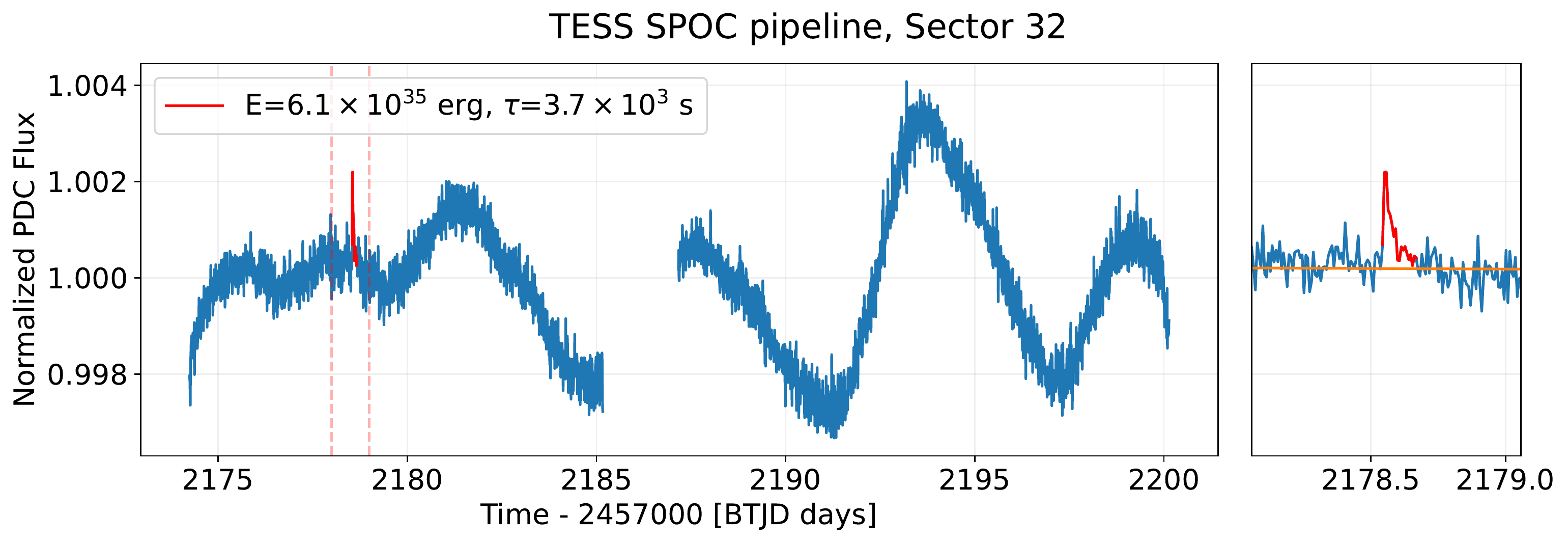}}
\caption{{\it TESS} light curves (blue lines) of HD\,36030 for two sectors when the object was observed. Visually selected flares in both sectors are highlighted in red and~also shown in the separate panels on the right. The~total energies of the flares as well as their durations are also shown. Orange lines represent the smooth interpolation of a quiescent light curve behaviour during the flare, used for removing background contribution from its energy. \label{fig:lc_tess}}
\end{figure}

In order to investigate the short time scale photometric variability of \HD, we used the data from the Transiting Exoplanet Survey Satellite (\tess; \citet{tess}).
\HD ~was observed by \tess\ twice during its ongoing operation, in~Sector~6 (11~December 2018--7 January 2019) of the prime mission, and~Sector~32 (19 November 2020--17~December 2020) of the extended mission. We downloaded the data products for these observations produced by the \tess\ Science Processing Operations Center pipeline (SPOC; \citet{tess_spoc}) and available at the Barbara A. Mikulski Archive for Space Telescopes (MAST)\endnote{\url{https://archive.stsci.edu}, accessed on March 10, 2023}. public data archive. We used the data products for the Long Cadence data, having 1800 s effective exposure during the prime mission, and~600 s---during the extended mission. 
In order to minimize the instrumental effects that are abundant in \tess\ data, we decided to use pre-search data conditioning simple aperture photometry (PDCSAP) light curves, which are corrected for instrumental trends using singular value decomposition, as~well as for flux contributions from nearby objects in crowded fields \citep{tess_spoc}. We also filtered out the measurements marked with bad quality flags. We did not apply any additional detrending or pre-processing to the data, and~the resulting light curves are shown in Figure~\ref{fig:lc_tess}.

The light curve from Sector~6 shows a strong flare that has already been reported by \citet{Maryeva2021flare} and \citet{Balona2021}. Visual inspection of Sector~32 also revealed a flare of smaller amplitude but similar duration. Both flares are shown in Figure~\ref{fig:lc_tess}. We characterized them by subtracting the smooth quiescent emission and integrating the flare profile to get its total fluence and fitted the fading part with an exponential function to get the characteristic duration, which we define as an e-folding~time.

We also estimated the total energetics of the flares by integrating the flare profile after subtraction of a smooth trend of a star' quiescent emission. In~order to correct for the fact that the majority of both the star and flare emission is outside of the {\it TESS} sensitivity range, we applied the bolometric corrections by convolving the spectrum of the star as well as flare spectrum with the sensitivity curve of {\it TESS}. As the temperatures of flares on the hot stars are not known yet, as~a conservative estimate,
following \mbox{\citet{Shibayama2013}} and \mbox{\citet{Gunther2020}}, we assumed the flares to have a blackbody spectrum with $T=9000$\,K, whereas for the star we used the spectral energy distribution of the best fitting {\sc Tlusty} model derived in Section~\ref{sec:tlusty}. Then we used these bolometric corrections to convert the relative amplitude of the flare to its relative total fluence, and,~ knowing the luminosity of \HD,\ we converted them to absolute values. 
The resulting flare energies are shown in Figure~\ref{fig:lc_tess}. 

\begin{figure} [H]

    \resizebox*{1.0\columnwidth}{!}{\includegraphics[angle=0]{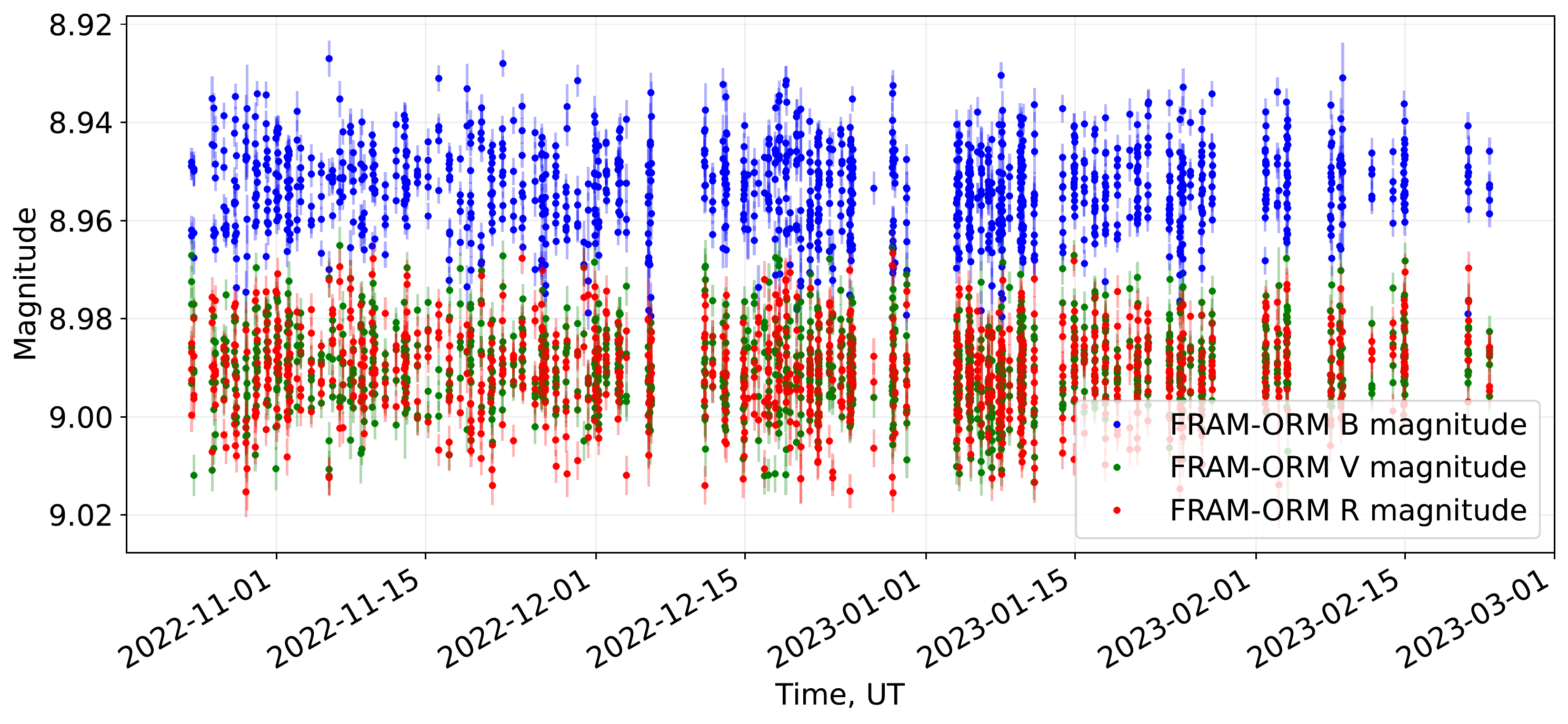}}
    \resizebox*{1.0\columnwidth}{!}{\includegraphics[angle=0]{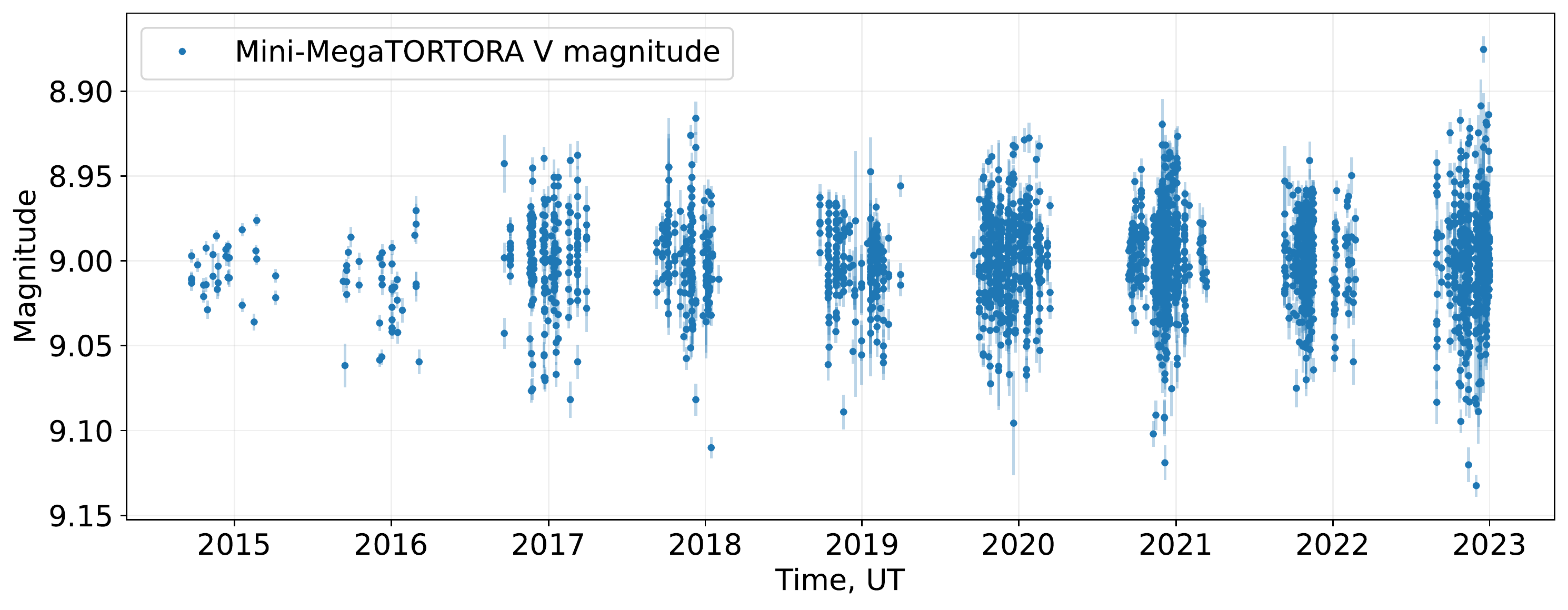}}

\caption{Light curves of \hd ~from FRAM-ORM (upper panel) and Mini-MegaTORTORA (lower panel). FRAM-ORM observed the star in three Johnson--Cousins photometric filters, whereas Mini-MegaTORTORA data are in white light and re-calibrated to Johnson $V$ bandpass. Both light curves lack any signs of systematic variability or statistically significant~flares.
 \label{fig:lc_fram}}
\end{figure}

In mid-October 2022, we also initiated a series of photometric observations of HD\,36030 on FRAM-ORM, which is a 10-inch Meade f/6.3 Schmidt--Cassegrain telescope with custom Moravian Instruments G2-1600 CCD installed in the Roque de Los Muchachos Observatory, La Palma. The~data were acquired in Johnson--Cousins $B$ (20 s exposures), $V$, and~$R$ filters (10 s exposures) to avoid saturation due to the brightness of the star and~then automatically processed by a dedicated Python pipeline based on the {\sc STDPipe} package \citep{stdpipe}, which includes bias and dark current subtraction, flat-fielding, cosmic ray removal, astrometric calibration, aperture (with a 5-pixel radius) photometry, and~photometric calibration using the catalog of synthetic photometry based on Gaia DR3 low-resolution XP spectra \citep{gaiadr3syn}. The~resulting light curve is shown in the upper panel of Figure~\ref{fig:lc_fram}.

Finally, in~order to assess the longer time scale variability of \hd, we acquired its photometry from the data archive of Mini-MegaTORTORA \citep{MMT2017}, which is a nine-channel wide-field optical monitoring system with high temporal resolution, operated since mid-2014 and located  at the Special Astrophysical Observatory, Nizhny Arkhyz, Russia. As~part of its systematic observations of the northern sky, primarily targeted at the detection and characterization of optical transients on a sub-second time scale, it acquires deeper ``survey'' images with 20 to 60~s exposures in white light, covering every point of the northern sky on average several times per night. These images are processed by a dedicated pipeline that, apart from standard calibration steps, determines the effective photometric system of every frame and~then employs this information to derive the $(B-V)$ colors of every star and re-calibrate the measurements to a Johnson $V$ filter \citep{Karpov2018}. The~resulting measurements are published online on the dedicated portal\endnote{\url{http://survey.favor2.info/}, accessed on March 10, 2023.}. The~data for \hd ~have been extracted from the Mini-MegaTORTORA archive and~passed through quality cuts in order to filter out the points corresponding to bad weather intervals and images where photometric calibration was too noisy. This resulted in more than 2300 points with good V magnitudes for the star, spanning more than 8 years since~mid-2014.

\begin{figure} [H]
    \resizebox*{1.0\columnwidth}{!}{\includegraphics[angle=0, trim=-0.3in 0 0 0]{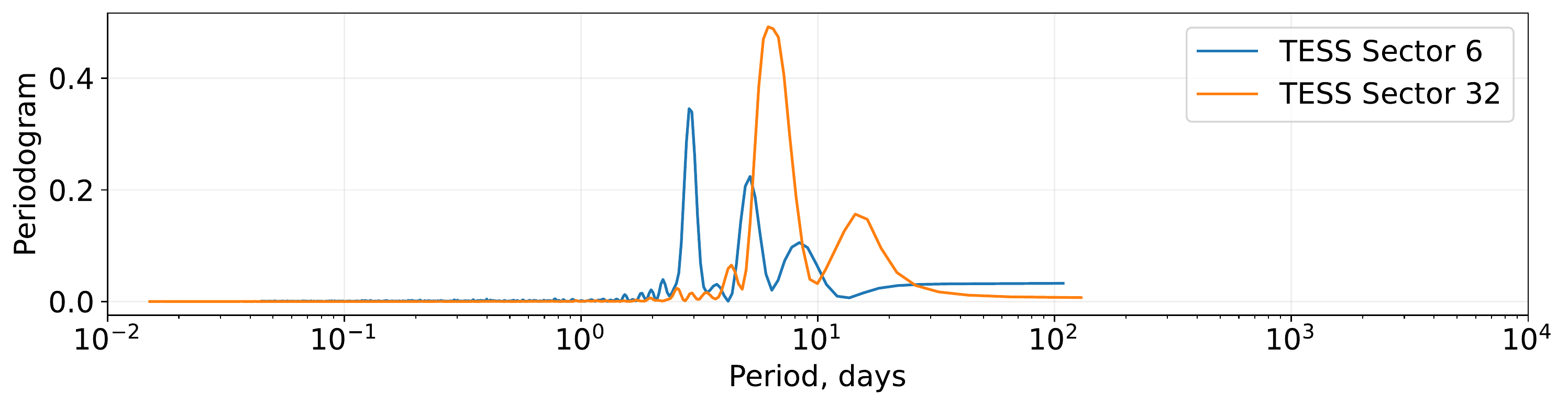}}
    \resizebox*{1.0\columnwidth}{!}{\includegraphics[angle=0]{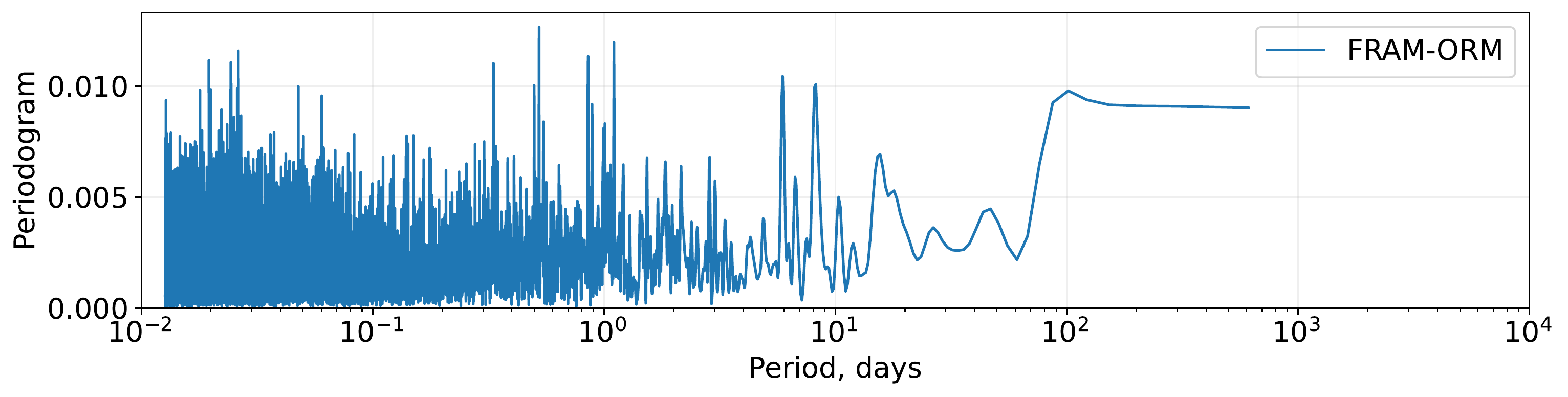}}
    \resizebox*{1.0\columnwidth}{!}{\includegraphics[angle=0]{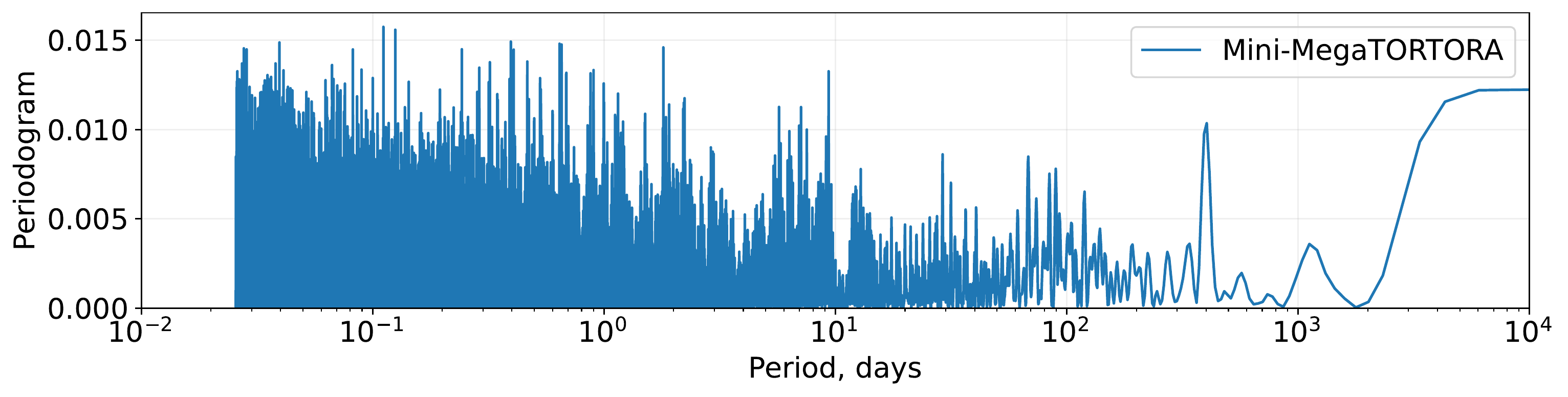}}
\caption{Lomb--Scargle periodograms of \hd ~light curves from {\it TESS}, FRAM-ORM, and Mini-MegaTORTORA. For~{\it TESS}, the~data for both sectors are shown in the upper panel. The~peaks there supposedly correspond to uncorrected instrumental effects in the light curves. The middle panel shows the multiband periodogram \citep{VanderPlas2015} for the FRAM-ORM light curve, and the lower panel shows the periodogram of Mini-MegaTORTORA data. There are no high-significance peaks visible in both of the latter panels. 
 \label{fig:lc_periodogram}}
\end{figure}

The light curves from both {\it TESS} sectors display prominent oscillating patterns, but~with sufficiently different amplitudes and characteristic time scales. The~upper panel of Figure~\ref{fig:lc_periodogram} shows the Lomb--Scargle periodogram \citep{Lomb1976,Scargle1982} of both light curves. There are no common peaks visible. Thus, we may attribute the oscillations to residual instrumental effects not fully corrected by the data processing, or~probably to some complex pulsational pattern of the star.
We also computed the periodograms of the FRAM-ORM and Mini-MegaTORTORA light curves. For~the former, we used the approach of \citet{VanderPlas2015} to get the combined periodogram of the measurements in all three colors. The~periodograms are also shown in Figure~\ref{fig:lc_periodogram}. They display no prominent peaks or structures common to all light curves, and~so we may conclude that we see no signs of any photometric periodicity in \hd.

\section{Discussion\label{sec:discussion}}

Our new spectroscopic data clearly display changes in the RV of \hd ~across different epochs of observations. It directly testifies to the binarity of the star, which is thus an SB1 binary, as~there are no spectral signatures of a second component in the spectrum. 
Although our data are too scarce to measure the orbital period, from~Figure~\ref{fig:radvel} we may constrain it as follows. The~fastest change we see it is about 10\kms\ between two consecutive nights, with~a total spread of velocities of about 38\kms. Thus, the~period should not exceed about 8 days, which gives the mass function of a second component $f = M_2^3 \sin^3i / (M_1 + M_2)^2 < 0.006 \Msun$. Thus, it favors a low-mass companion on an orbit with low to moderate inclination, which is consistent with the absence of eclipses in the light curve. The~semi-major axis of the orbit is $a<1.7\cdot10^{12}$ cm assuming the total system mass of $3\Msun$ and the period shorter than 8 days.  Further spectral monitoring is necessary in order to better constrain the parameters of the second component of the system, as~well as orbital~parameters.

On the other hand, we do not see any signs of periodic variations in the light curves (see Figures~\ref{fig:lc_tess} and \ref{fig:lc_periodogram}) on these time scales, apart from quasi-periodic structures in the {\it TESS} data, with both periods and amplitudes varying both within the spans of individual observations and between different epochs, which we cannot attribute to the binary period. Their frequencies are also lower than the ones of rotational modulation seen in B-type stars in both {\it Kepler} \citep{Balona2016} and {\it TESS} \citep{Balona2019} data, so we assume that these oscillations are also not a signature of stellar rotation that might appear due to, e.g.,~starspots on the surface of HD\,36030, but~most probably they are a signature of uncorrected systematic effects in the data (see, e.g.,~\citet{Hattori_2022}).

\begin{figure} [H]
\resizebox*{1.0\columnwidth}{!}{\includegraphics[angle=0]{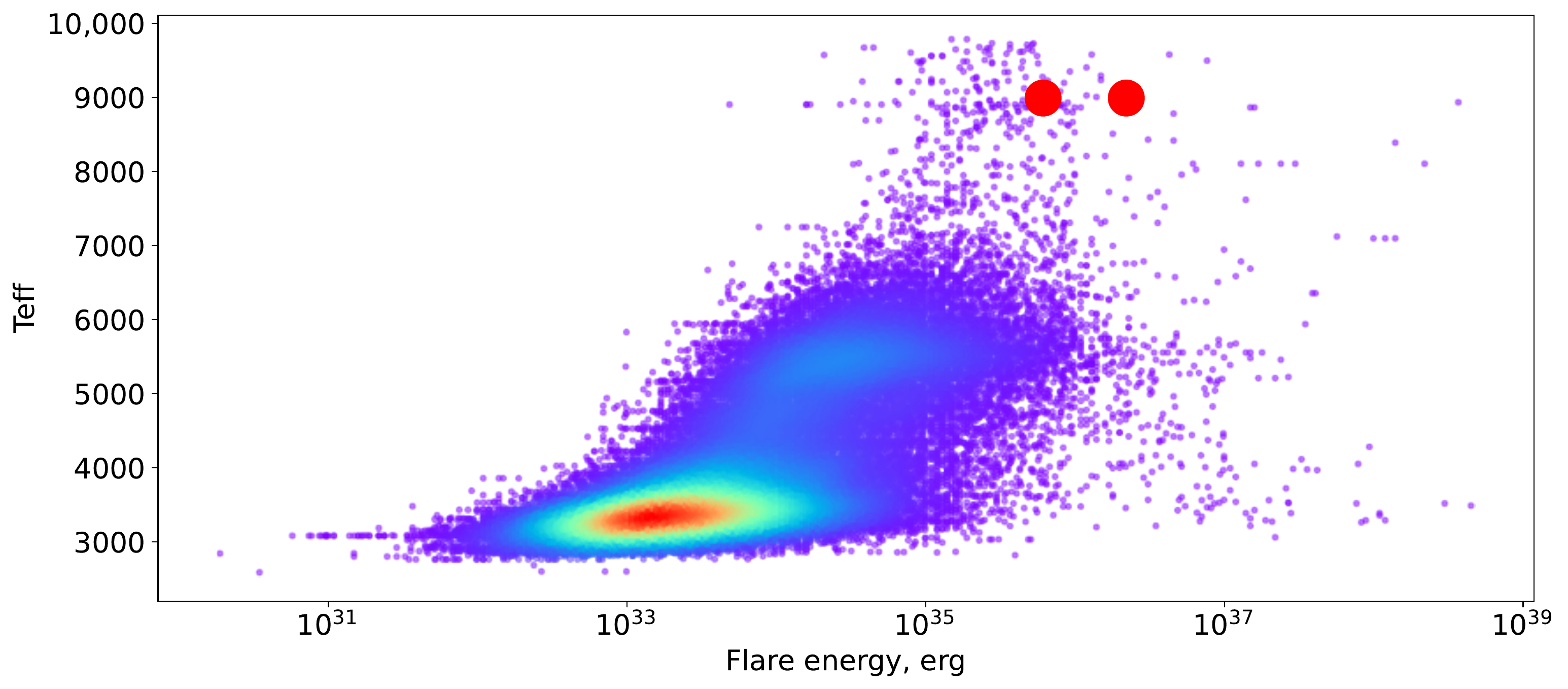}}
\caption{Energies of the flares of \HD ~(red circles) in comparison with the ones detected by \citet{Yang2023}, as~a function of the effective temperatures of the stars. The~temperatures are originating from Gaia~DR2 \citep{gaiadr2} and are underestimated for the hottest objects. The~colors of the dots represent the density of points.
 \label{fig:energies}}
\end{figure}

We detected two flares on \HD ~in the {\it TESS} data from two sectors, i.e.,~one in each 27-day series of continuous observations separated by nearly two years, meaning that these flares are not some extreme events but occur regularly. It is consistent with the detection of repeating flares on hot stars by \citet{Yang2023}. The~energies of these flares\endnote{We must note that our estimation of flare energies is a conservative one, as it assumes their temperatures to be 9000 K, which may not be a good approximation, as the effective temperature of the star itself is above that. Moreover, there are signatures that the temperatures of superflares even on cool stars may be significantly larger than that \citep{Howard2020}. Thus, the~actual energies of the flares we detected may also be significantly larger.}, $2.2\cdot10^{36}$ and $6.1\cdot10^{35}$ erg, place them among the most energetic ones detected by \citet{Balona2021} and \citet{Yang2023} (see Figure~\ref{fig:energies}). This fact makes it highly improbable that the flares originate from the low-mass companion\endnote{The same argument also holds against any other similar source in the field of \HD\ ~that may occasionally pollute the large aperture of \tess.} of \HD, which may only be a late-type, low-luminosity star due to the absence of any spectroscopic signature  from it and~a very low value of its mass function, which implies its low mass for reasonable range of values of the system' inclination angle. On~the other hand, our modeling of \HD ~suggests that it lacks the convective layers that may produce magnetic fields. 
 There are no conclusive signs of surface magnetic fields---no strict periodicity in the light curve disfavors the presence of starspots; however, quasi-periodic patterns seen in Figure~\ref{fig:lc_tess}, if~they are not instrumental in nature, may in principle be the manifestation of short-lived spots on \HD\ surface---but in the absence of convection, the formation of such spots may be unrealistic.
The star also lacks any signs of peculiarities in its spectra, such as additional emission components in hydrogen lines that might be a sign of a circumstellar outflow or disk  that might help generate the magnetic fields due to star--disk interaction. 
However, the~presence of a low-mass companion star, most probably possessing significant magnetic fields, on~a close orbit around \HD\ may be the source of magnetic fields spanning across the system and both storing the energy enough for powering the flares and~maybe inducing the starspots on \HD\ surface.

Thus, we may conclude that the only mechanism that may produce the magnetic fields necessary for powering the energetic flares may be the interaction between components inside the binary~system.

\section{Conclusions \label{sec:conclusions}}

We performed a dedicated study of \HD, which was the hottest flare star detected by \citet{Maryeva2021flare}. We found one more flare in the {\it TESS} data that confirmed the repeating nature of the flaring. We initiated a spectroscopic monitoring of the star in order to better understand the physics behind these flares. The~spectra did not display any peculiarities and suggest that \HD ~is a normal main-sequence B9\,V star. On~the other hand, we clearly detected radial velocity changes between different epochs of observations, thus confirming the binarity of the star. 
We did not detect any coherent variability in the light curve of the star, so we could not estimate the period of the orbital motion of the star or~confirm the presence of the spots on its surface. The~latter may be a sign of the absence of strong surface magnetic fields on HD\,36030. Thus, the~question of the origin of magnetic field powering the strong flares from \HD\ is still open, and~we favor the magnetic interaction with a second low-mass component in a binary system as their cause.

\vspace{6pt} 

\authorcontributions{
O.M. proposed the concept of the study, performed observations on the Perek 2-m telescope, reduced the spectroscopic material, and performed spectral analysis; P.N. performed  numerical modeling of the stellar atmosphere using {\sc~Tlusty} code; 
S.K. performed photometric monitoring, reduction, and~analysis of photometric data, and~estimated the properties of flares. 
All authors participated in the discussion of the results and preparation of the~manuscript. All authors have read and agreed to the published version of the manuscript.
}

\funding{This research received funding from the European Union's Framework Programme for Research and Innovation Horizon 2020 (2014--2020) under the Marie Sk\l{}odowska-Curie Grant Agreement No. 823734  (POEMS project). The work based on data taken with the Perek telescope at the Astronomical Institute of the Czech Academy of Sciences in Ond\v{r}ejov, which is supported by the project RVO:67985815  of the Academy of Sciences of the Czech Republic.
P.N. acknowledges support from the Grant Agency of
the Czech Republic (GA\v{C}R 22-34467S) and from the Polish National
Science Centre under projects No. UMO-2017/26/E/ST9/00703 and
UMO-2017/25/B/ST9/02218. 
S.K. acknowledges support from the European Structural and Investment Fund
and the Czech Ministry of Education, Youth and Sports (Project CoGraDS---CZ.02.1.01/0.0/0.0/15\_003/0000437).
This research has used the services of \mbox{\url{www.Astroserver.org}}. 
The operation of the robotic telescope FRAM-ORM is supported by the grant of the Ministry of Education of the Czech Republic LM2018102. 
The operation of the Mini-MegaTORTORA was supported  under  the   Ministry of Science and Higher Education of the Russian Federation grant  075-15-2022-262 (13.MNPMU.21.0003). 
This paper includes data collected by the {\it TESS} mission, which are publicly available from the MikulskiArchive for Space Telescopes (MAST). Funding for the {\it TESS} mission is provided by NASA's Science Mission directorate. This research was made by using of the SIMBAD database and the VizieR catalogue access tool, both operated at CDS, Strasbourg, France. 
}

\institutionalreview{Not applicable.}

\informedconsent{Not applicable.}

\dataavailability{The data presented in this study are available on request from the corresponding author, or available in publicly accessible data archives as specified in the manuscript text.} 

\acknowledgments{The authors acknowledge the help from Suryani Guha with observations with Coud\'e spectrograph at the Perek 2 m~telescope.  }

\conflictsofinterest{The authors declare no conflict of~interest.} 

\begin{adjustwidth}{-\extralength}{0cm}
\printendnotes[custom] 

\reftitle{References}

\PublishersNote{}
\end{adjustwidth}
\end{document}